\documentclass[a4paper,12pt]{article}

\usepackage{natbib}

\usepackage{soul,xcolor}
\setstcolor{black}

\usepackage{sectsty}

\allsectionsfont{\normalsize\bfseries}

\allsectionsfont{\centering}

\usepackage{graphicx}

\usepackage{multirow}

\usepackage{hhline}

\usepackage{times}

\usepackage{lipsum}

\usepackage{wrapfig}

\usepackage{titling}

\let\OLDthebibliography\thebibliography

\renewcommand\thebibliography[1]{

  \OLDthebibliography{#1}

  \setlength{\parskip}{0pt}

  \setlength{\itemsep}{0pt plus 0.3ex}

}

\begin{document}

{\hskip -0.6cm

\begin{minipage}[h]{16cm}


\textbf{Deterministic chaos in the X-Ray sources}

\vspace{\baselineskip}

        M.~Grzedzielski$^{1*}$,

        P.~Sukova$^{1}$,

        and~A.~Janiuk$^{1}$

        \vspace{\baselineskip}

$^{1}$ \textit{Center for Theoretical Physics, Polish Academy of Sciences, Al. Lotnik\'ow 32/46 02-668 Warsaw, Poland}\\


\vspace{\baselineskip}


\end{minipage}

}

{

\hoffset -3.6cm

\hskip 1.35cm

\begin{minipage}[h]{12cm}

{\textbf{Abstract.} \textit{Hardly any of the observed black hole accretion disks in X-Ray binaries and active galaxies shows constant flux. When the local stochastic 
 variations of the disk occur at specific regions where a resonant behaviour takes place,
there appear the  Quasi-Periodic Oscillations (QPOs). If the global structure of the flow and its non-linear hydrodynamics affects the fluctuations,
the variability is chaotic in the sense of deterministic chaos.
 Our aim is to solve a problem of the stochastic versus deterministic nature of the black hole binaries variability.
 We use both observational and analytic methods.   
We use the recurrence analysis and we study the occurence of long diagonal lines in the recurrence plot 
of  observed data series and compare it to the surrogate series.  
 We analyze here the data of two X-Ray binaries - XTE J1550-564,
 and GX 339-4 observed by Rossi X-ray Timing Explorer. In these sources, the non-linear variability 
 is expected because of the global conditions (such as the mean accretion rate) leading to the possible instability 
 of an accretion disk. The thermal-viscous instability and fluctuations around the fixed-point solution occurs 
 at high accretion rate, when the radiation pressure gives dominant contribution to the stress tensor.
 }
}
 

 \end{minipage}

}











%



\section{Introduction}

\label{intro}
Appearance of Quasi Periodic Oscillations (QPOs) may prove the evolution of the accretion disk governed by the
limit cycle oscillations due to the radiation presure instability.
The classic theory of accretion as proposed by \citet{shakura} and \citet{lightman} predicts that the disk is unstable and undergoes the limit cycle oscillation, if the viscous
stress tensor scales with the total (gas plus radiation) pressure and the global accretion rate is large enough for the radiation
pressure to dominate. The accretion disks may undergo the limit-cycle 
oscillations around a fixed point due to the two main types of thermal-viscous instabilities - radiation pressure instability and ionization instability. Both these instability types are known for 
over 40 years in theoretical astrophysics. In classical theory of \citet{shakura}, the accretion flow structure is based on $\alpha$ description for the 
viscous energy dissipation.  It assumes that 
the non-zero component $T_{r \phi}$ of the stress tensor is proportional to the total pressure. The latter includes the radiation pressure
which scales with temperature as $T^4$ and blows up in hot disks for large accretion rates. This in turn affects the heating and cooling balance between the energy dissipation and radiative losses. If the accretion rate is small, then most
of the disk is gas pressure dominated and stable. For large enough accretion rates, there appears a zone where some of its annuli are dominated by radiation pressure and unstable.
 The X-Ray binaries, such as GRS1915 or IGR J17091, present many states of variability \citep{belloni}.
 The Quasi-Periodic Oscillations with heartbeat patterns usually occur in disk-dominated soft or soft-intermediate states \citep{nandi,janiuk2015}. 
 We 
 investigate the behaviour of time series of two other sources
 (GX 339-4 and XTE J1550-564) to find the tracks of deterministic 
chaos to answer the question - are the luminosity oscillations driven by deterministic or stochastic system.
For that purpose we employ the recurrence analysis and we compare the results 
with the surrogate data.
Here we pursue a similar analysis as \citet{misra} to other sources, but
we use a novel method which is the recurrence analysis.
\section{Recurrence analysis}
To answer the question about the dynamics of the system - is it stochastic or deterministic, we need a method, 
which reconstructs the multidimensional
phase space, and investigates its behaviour.
We use the recurrence analysis, tool used to study the time series \citep{eckmann}. 
 The recurrence analysis works in a broad range of application. 
We combine this method with the surrogate data approach developped by \citet{theiler} in the following way.
We first pose the ``null hypothesis'' about the measured time series, that the data are product of temporally independent white noise or linearly autocorrelated gaussian noise.
We construct surrogates using the publicly available software package software package TISEAN \citep{Schreiber200346,1999chao.dyn.10005H} and 
apply the recurrence analysis  \citep{marwan}.
We define significance  $\bar{\mathcal{S}}$ of the
non-linear behaviour as a weighted difference between
the estimate of R\' enyi's entropy for the observed 
data $K_2^{\rm obs}$ and the ensamble of its surrogates
\citep{sukova}.

The basic object of the analysis is the recurrence matrix, which describes the times, when the trajectory returns close to itself (closer than certain threshold $\epsilon$). 
The recurrence matrix is defined as follows:
\begin{equation}
\mathbf{R}_{i,j}(\epsilon) = \Theta (\epsilon - \parallel \vec{y}_i - \vec{y}_j \parallel ), \qquad i,j = 1,...,N, \label{RP_def}
\end{equation}
where $\vec{y}_i = \vec{y}(t_i)$ are ($N$) points of the reconstructed phase trajectory and $\Theta$ is 
the Heaviside step function.

The recurrence plot (RP) is a visualisation of the recurrence matrix, in which the matrix
elements equalled to 1 are plotted as black dots. In such plot the long diagonal lines corresponds
to the case, that the trajectory behaves similarly in two different times. Hence RP of a regular 
trajectory would consists of a set of diagonal lines only, contrary to the randomly scattered points 
in RP of stochastic system.

R\' enyi's entropy $K_2$ is related with the cumulative histogram of diagonal lines $p_c(\epsilon,l)$,
describing the probability of finding a line of minimal length $l$ in the RP, by the relation
\begin{equation}
p_c(\epsilon,l) \sim f(\epsilon) e^{-l \Delta t K_2} \nonumber, \label{cumul}
\end{equation}where $f(\epsilon)$ is a known function of $\epsilon$, we can estimate the value of $K_2$ as the slope
of the logarithm of the cumulative histogram versus $l$ for constant $\epsilon$.
Because measured data do not provide the phase space trajectory, it has to be reconstructed from the
observed time series with the time delay technique. 
The resulting phase space vector is given as
\begin{equation}
\vec{y}(t) = \{x(t),x(t+\Delta t),x(t+2\Delta t),\dots,x(t+(m-1)\Delta t)\}, 
\end{equation}
where $x(t)$ is the time series, $\Delta t$ is the embedding delay and $m$ 
is the embedding dimension. \\   We extract the lightcurve for every observation, rescale it to zero mean and zero variance.
 We find appropriate guess of the embedding dimension. We produce 100 surrogates for 
 the different recurrence threshold $\epsilon$. For each recurrence threshold we compute the
 estimator of Renyi entropy. Later we compute the average significance in respect to 
 the surrogates. The Renyi entropy of the order $\alpha$ is defined as follows:
 \begin{equation}
  K_\alpha = \frac{1}{1-\alpha} \ln \sum\limits_{i=1}^{i=N} p_i^\alpha,
 \end{equation}where $p_i$ is the probability that the random variable has a value of $i$. It describes the randomness of the system.
 The definition of significance of chaotic process, depending on 
 threshold $\epsilon$  is following:
\begin{equation}
  \mathcal{S}(\epsilon) = \frac{N_{sl}}{N_{\rm surr}} \mathcal{S}_{sl}- s \frac{N_{\mathcal{S}_{K} }}{N^{\rm surr} } \mathcal{S}_{K_2} (\epsilon). 
 \end{equation}
 $s=$sign$(\mathcal{Q}^{obs}(\epsilon)- \bar{\mathcal{Q}}^{\rm surr} (\epsilon))$, $N^{\rm surr}$ is the total number of surrogates and $N_{\mathcal{S}_K}$ is the number of surrogates 
 which have enough long diagonal 
lines for the estimate of $K_2$, so that $N_{\rm sl} + N_{\mathcal{S}_K} = N^{\rm surr}$. 
 $\mathcal{Q}^{obs}$ and $\bar{\mathcal{Q}}^{\rm surr}$ are the natural logarithms of $K_2$ for observed and surrogate data.
 $\mathcal{S}^{sl}=3$ and $\mathcal{S}_{K_2}$ is the significance computed only from the surrogates, which have enough long lines according
 to relation with respect to the standard deviation of $\bar{\mathcal{Q}}^{\rm surr}(\epsilon)$, computed for set of surrogates.
 The standard deviation is denoted as $\sigma_{\bar{\mathcal{Q}}^{\rm surr}(\epsilon) }$ 
  \begin{equation}
  \mathcal{S}_{K_2} (\epsilon)  = \frac{|\mathcal{Q}^{obs}(\epsilon) - \bar{\mathcal{Q}}^{\rm surr}(\epsilon) |}{\sigma_{\bar{\mathcal{Q}}^{\rm surr}(\epsilon) }}.
 \end{equation}
 \\
\section{Observations and results}
\subsection{Observations}
We obtained some RXTE PCA observations of GX339-4 and XTE J1550-564 with small binning time.
We extract the lightcurves using Heasoft 6.16 high energy astrophysics software package. We adjust the proper binning time to minimize the error and simultanously
to not lose the information about oscillations at the scale of several seconds or several tens of
seconds.

 In Figure 1, we present lightcurve of GX 339-4. The data were extracted using \textit{sefilter} in \textit{General Event} mode
from channels 0-24 of RXTE PCA (2-10 keV). The binning was 0.5s.

 In Figure 2, we present lightcurve of  XTE J1550-564. The data were extracted using \textit{saextrct}
 in \textit{standard1} mode. The binning was 0.125s. Both Figures 
1 and 2 present lightcurves with characteristic variability pattern, that resembles the so-called ``heartbeat
oscillations'' already known for the sources IGR J17091-3624 and GRS 1915+105 \citep{Altamirano et al. (2011),
belloni}.

 \subsection{Results}
 
 We analyse these observations using the recurrence analysis, which can give us the important hint for existence of deterministic
 chaos in the time series.
 
  In Figure 3 we show length of the longest line in the recurrence plot $L_{max}$.
  
 The length of the longest line on the recurrence plot is a hint for deterministic and regular character of the probed
sequence (lightcurve). The most regular sequences have the longest lines spreading across all the Figure, otherwise if we have 
stochastic sequence like the surrogate data, there exist the values of $\epsilon$ for which the longest line is short in comparison
to the longest line in the sequence with the same Power Density Spectrum, but with greater regularity.

The Figure 3 describes the times, when the time series trajectory returns to itself closer the than $\epsilon$. The strong
difference between the real data and surrogate series is seen and $L^{max}_{obs} > L^{max}_{\rm surr}$ for a wide range of $\epsilon$. This result indicates non-linear 
features in the dynamics of the source GX 339-4.
 
 Figure 4 was made for the data of the source XTE J1550-564. It shows, that the significance is dependent on the recurrence threshold $\epsilon$, and
is usually decreasing for very high $\epsilon$. In our analysis we consider the averaged significance, which is represented in the plot by the red horizontal line. 
For details about the computations of the significance see \citet{sukova}.

Apart from $\epsilon$, the significance also depends on the other
parameters used for computing the recurrence matrix, like embedding delay $\Delta t$ and embedding dimension $m$.

As was shown by \citet{thiel} , 
the estimation of dynamical invariants like Renyi entropy $K_2$ should not 
depend on these parameters for a
trajectory of a particular dynamical system. However, we are dealing with
real observational data instead of ideal time series of certain dynamical system.

These data are contaminated both by the uncertainties emanating from the observing
instrument and also by the stochastic nature of thermal radiation, which is 
the underlying physical process emitting the observed flux. Hence in our case,
the significance depends also on these parameters. In Fig. 5 
you can see the dependence of significance for the observation from Fig 1 
on the used embedding dimension. The dependence shows a plateau of almost
constant values in the range of $m \in (3,6)$, decreasing slightly for $m \in (7,9)$,
but it has a sharp decline for m=10.

We presume, that this behaviour is due to the noise contained 
in the data and that the strength of the dependence is induced
by the strength of the noise in the data. For comparison we show also Figure 6
computed for the observation from Figure 2. In this case, the decrease of the significance with embedding dimension 
is much slighter. 
\begin{figure}
\includegraphics[width=0.95\textwidth]{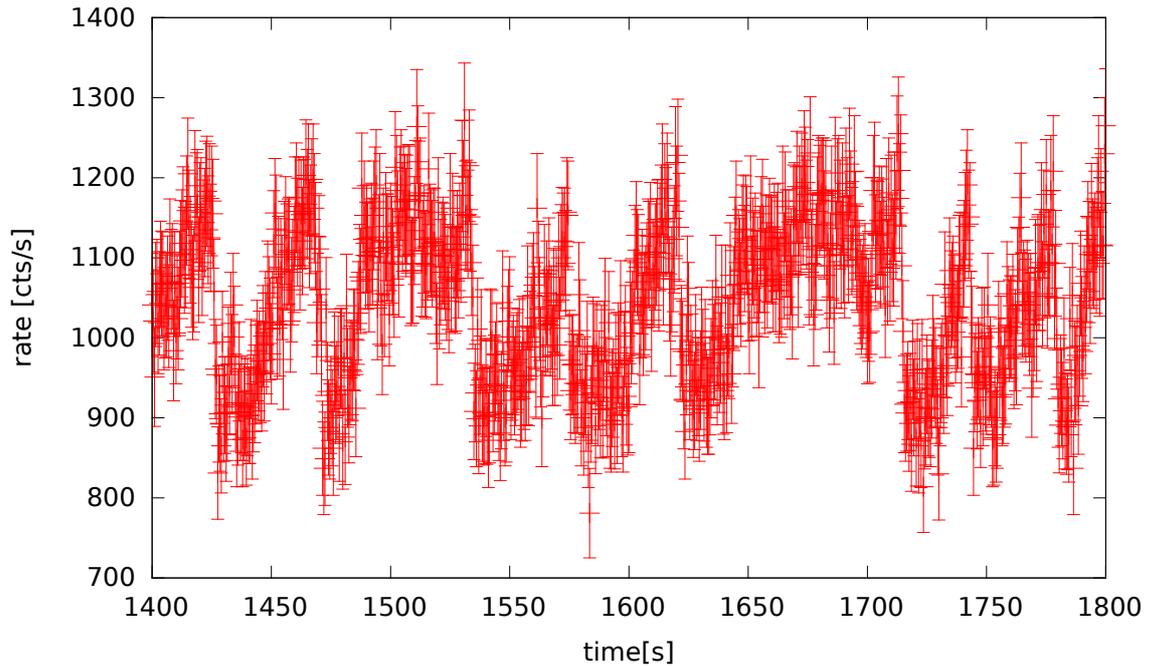}
\caption{Lightcurve of GX 339-4 on 29.04.2010 in Soft-Intermediate State \citep{nandi}, 
ObsID 95409-01-16-05, extracted from PCU2 with time bin 0.5s for the energy range 2-10 keV (channels 0-24).}
\end{figure}
\begin{figure}
\includegraphics[width=0.95\textwidth]{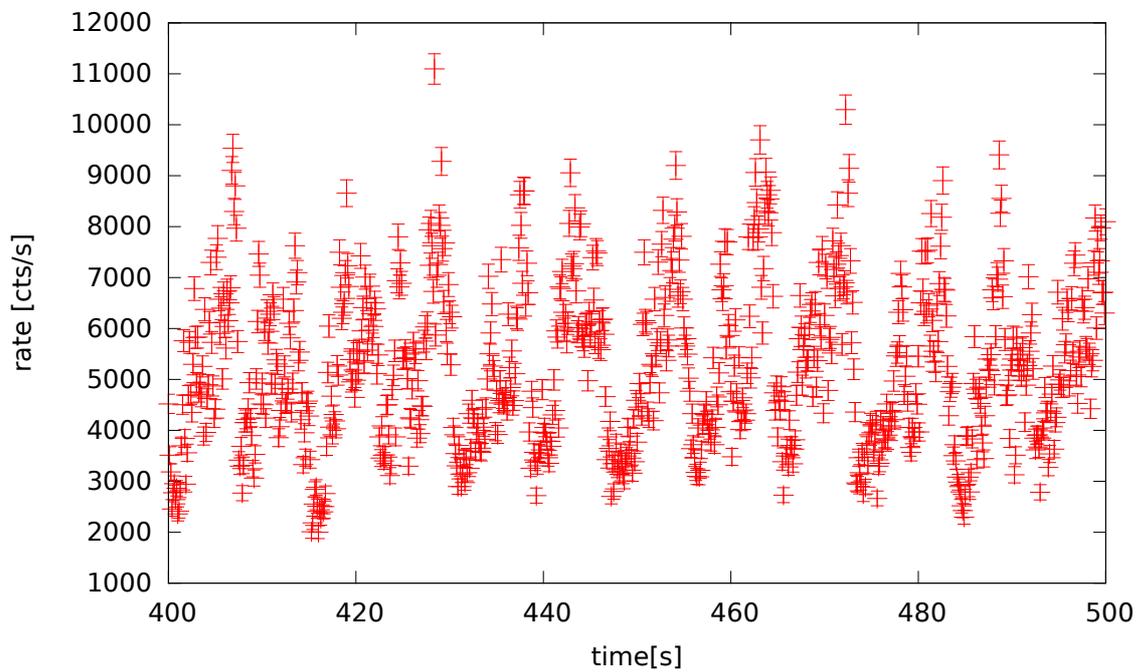}
\caption{Lightcurve of XTE J1550-564 on 08.09.1998, ObsID 30188-06-03-00, extracted with time bin 0.125s
from Standard 1 data mode (2-60 keV). The same observation was studied by \citet{sobczak}.}
 \end{figure}
\begin{figure}
\includegraphics[width=0.99\textwidth]{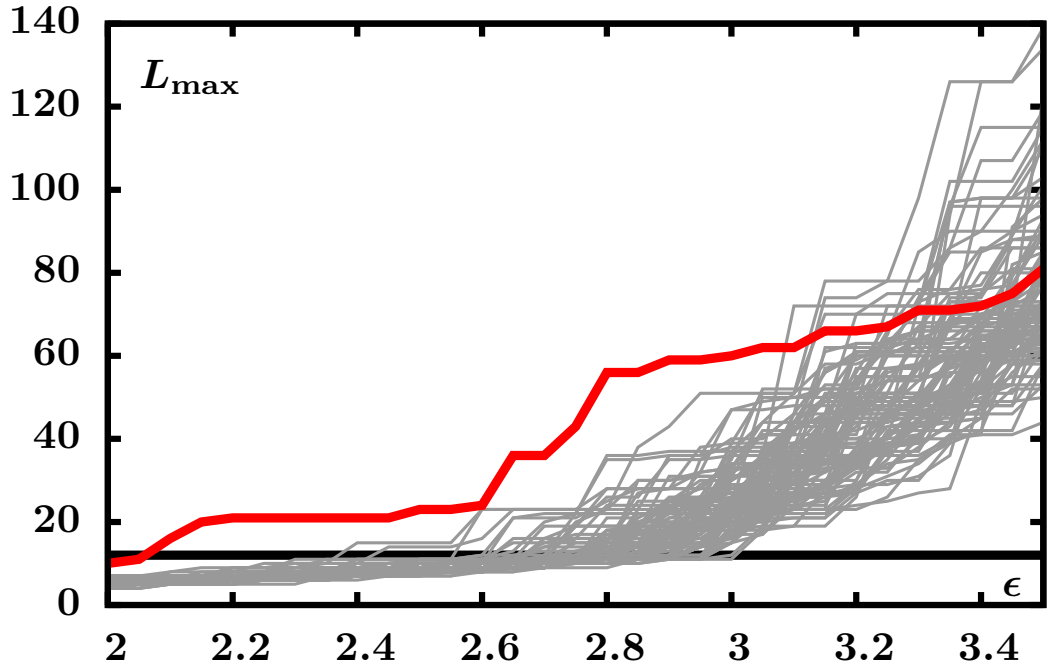}
  \caption{  Dependence of $L_{\rm max}$ on $\epsilon$ for the observation shown on Fig. 1  by thick
  red line and the ensemble of 100 surrogates by grey thin lines. The length of the
  longest line is expressed in the number of
  points on the line. 
  For some values of $\epsilon$ - for example for the interval 
  $\epsilon \in [2.2,3]$ it clearly shows, that the length 
  of the longest line for the observation is significantly different from all ensemble of
  surrogates. It means that there exist the tracks of nonlinear dynamics for that observation of
  the microquasar GX 339-4.
}
\end{figure}
\begin{figure}
\includegraphics[width=0.99\textwidth]{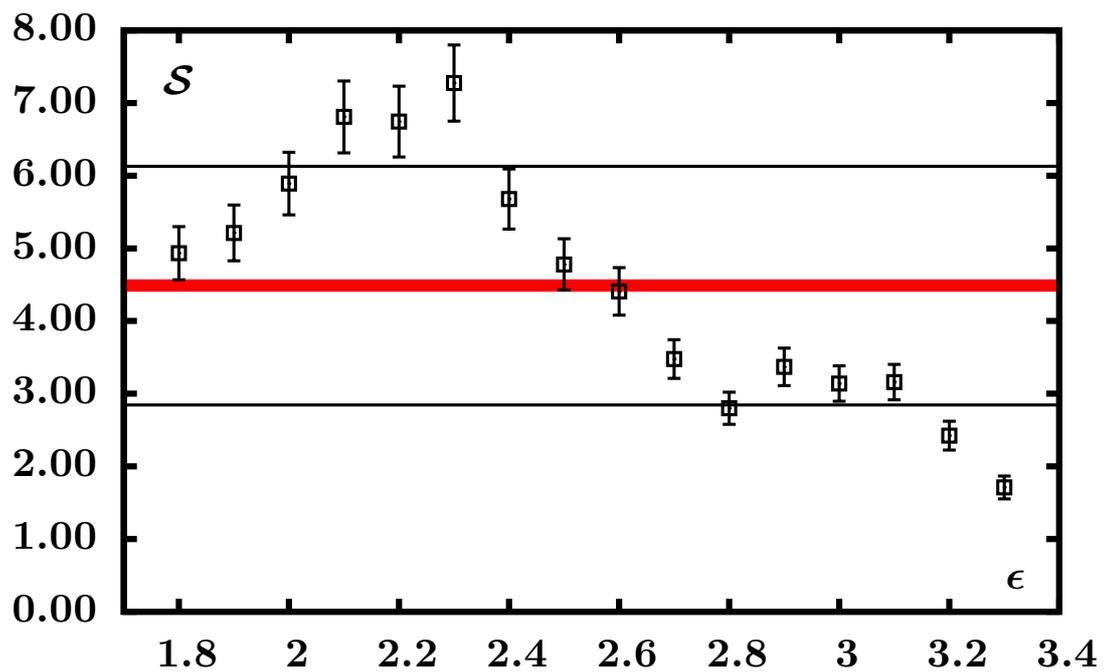}
\caption{  
Dependence of significance of the non-linear nature of the observation of  XTE J1550-564 
on 28.09.1998, ObsID 30191-01-14-00 on
the recurrence threshold $\epsilon$ and the mean (red horizontal line) and standard deviation 
(black horizontal lines) of the values
of significance.}
\end{figure} 
\begin{figure}
\includegraphics[width=0.99\textwidth]{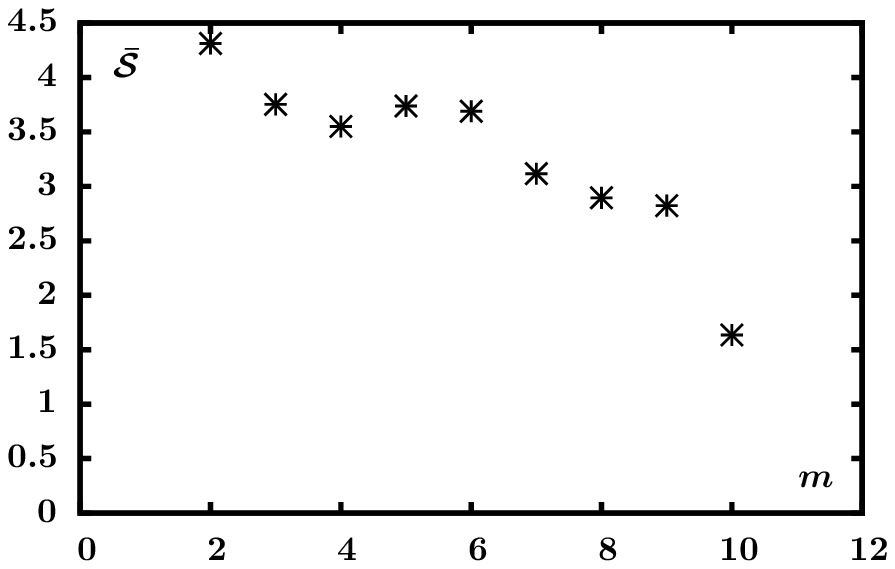}
\caption{  
Dependence of significance of the non-linear 
nature of the observation from Figure 1 on
embedding dimension $m$.}
\end{figure} 

\begin{figure}
\includegraphics[width=0.99\textwidth]{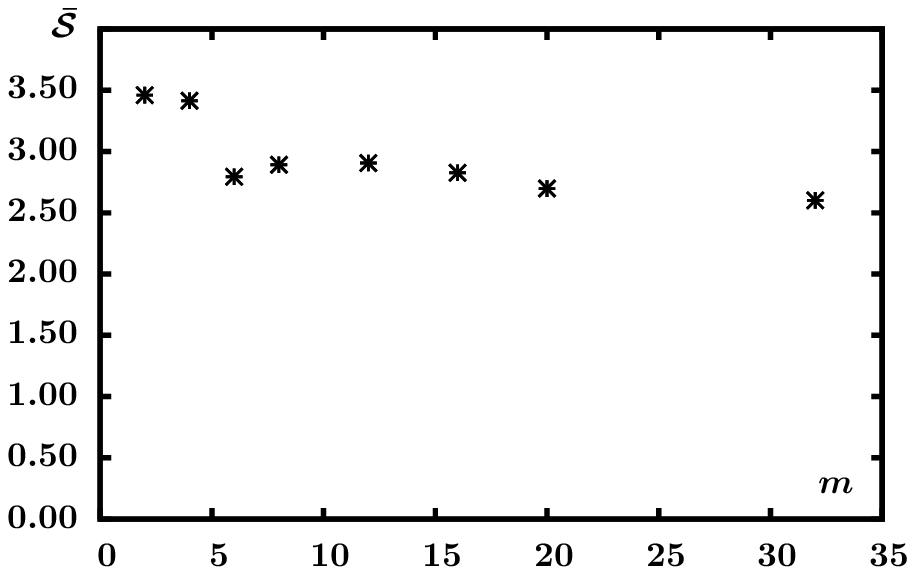}
\caption{  ependence of significance of the non-linear
nature of the observation from Figure 2 on embedding dimension $m$.}
\end{figure}

\section{Discussion and conclusions}
\subsection{Discussion}
The high energy radiation emitted by black hole X-ray binaries originates in an accretion disk.
Most of the sources undergo fast and complicated variability patterns on different timescales.
The variations that are purely stochastic in their nature, are expected since the viscosity of the accretion disk is connected with
its turbulent behaviour induced by magnetic instabilities.
The variability of the disk that reflects its global evolution governed by
the nonlinear differential equations of hydrodynamics, may not be only 
purely stochastic. Instead, if the global conditions in
 the accretion flow are such that the system finds itself in an unstable
configuration, the large amplitude fluctuations around the fixed point
solution will be induced. The observed behaviour of the disk will
then be characterized by the deterministic chaos.
 The recent hydrodynamical simulations of the global accretion disk 
evolution confirm that the quasi-periodic flare-like events observed in 
couple sources, are in a good quantitative agreement with the radiation pressure instability model \citep{janiuk2015}.

At least eight of the known BH X-ray binaries should have their 
Eddington accretion rates large enough for the radiation pressure 
instability to develop.

  We used the recurrence analysis method to study the non-linear behaviour of several X-ray sources mentioned in
  \citet {janiukczerny}.
  Our present analysis confirmed that the
variability in these sources is significantly governed by the nonlinear dynamics of accretion process.
We confirm that the GRS 1915 and IGR J17091, which show deterministic chaos, and are not the only sources 
with that feature.


 XTE J1550-564 is classified as a microquasar on the basis of its large-scaled moving jets, detected at X-ray and
 radio wavelengths \citep{Corbel et al. (2002)},
similarly with GX 339-4 \citep{corfe}.
We expect that these two objects should have similar characteristics of the disk variability, and along with the two
well studied microquasars the non-linear dynamical processes in 
 these sources should also occur. Our current analysis
does confirm these expectations.

\subsection{Conclusions}
We applied the recurrence analysis on observations of 
two black hole X-ray binaries observed by RXTE satellite.
We developed a method for distinguishing between stochastic, non-stochastic linear and
non-linear processes using the comparison of the quantification of recurrence plots with the
surrogate data. We tested our method on the sample of observations of the microquasar IGR J17091-3624,
which spectral states were
provided by \citet{Pahari et al. (2014)}. Significant results for the heartbeat state were obtained.
 We examined several observations of the other 
 microquasars.
Aside from the well-studied binaries GRS 1915+105 and 
IGR J17091-3624, we found significant traces of non-linear dynamics also in
three other sources:
GX 339-4, XTE J1550-564 as discussed in this
article, and GRO J1655-40, as presented in \citet{sukova}.
Further details can be found in \citet{sukova}.
\section{Acknowledgements}
This work was supported in part by the grant DEC-2012/05/E/ST9/03914 from the Polish National Science Center.
\bibliographystyle{apa}

\end{document}